\documentclass[prb, twocolumn,preprintnumbers,amsmath,amssymb]{revtex4}

\newcommand{\m}[1]
{\mathrm{#1}}

\usepackage{graphicx}
\usepackage{dcolumn}
\usepackage{bm}
\usepackage{times}

\begin{document}

\bibliographystyle{unsrt}

\title{Absorptive and dispersive optical responses of excitons in a single quantum dot}

\author{Benito Al\'en$^{\rm 1}$\footnote[0]{\scriptsize$^{\rm 1)}$Present affiliation: Instituto de Microelectr\'onica de Madrid, CSIC-CNM Parque Tecnol\'ogico de Madrid, Isaac Newton 8, 28760 Tres Cantos, Madrid, Spain}}
\author{Alexander H\"{o}gele$^{\rm 2,}$\footnote[0]{\scriptsize$^{\rm 2)}$Present affiliation: Institute of Quantum Electronics, ETH H\"{o}nggerberg HPT G10, CH-8093 Z\"{u}rich, Switzerland.}$^{\rm *}$\footnote[0]{\scriptsize$^{\rm *)}$Corresponding author. E-mail: hogele@phys.ethz.ch}}
\author{Martin Kroner}
\author{Stefan Seidl}
\author{Khaled Karrai}
\affiliation{Center for NanoScience, Department f\"{u}r Physik,
Ludwig-Maximilians-Universit\"{a}t, Geschwister-Scholl-Platz 1,
D-80539 M\"{u}nchen, Germany}

\author{Richard J. Warburton}
\affiliation{School of Engineering and Physical Sciences,
Heriot-Watt University, Edinburgh EH14 4AS, United Kingdom}

\author{Antonio Badolato$^{\rm 2}$}
\author{Gilberto Medeiros-Ribeiro}
\author{Pierre M. Petroff}
\affiliation{Electrical and Computer Engineering Department, University of California, Santa Barbara, California 93106, USA}

\date{\today}

\begin{abstract}
We have determined both the real and imaginary parts of the dielectric polarizability of a single quantum dot. The experiment is based on the observation and the manipulation of Rayleigh scattering at photon frequencies near the resonance of an optical exciton transition in single self-assembled InAs and InGaAs quantum dots. The interference between the narrow-band laser field and the weak electromagnetic field coherently scattered by the quantum dot is detected with a cryogenic Fabry-P\'{e}rot setup by combined differential transmission and reflectivity measurements.
\end{abstract}


\maketitle

Considerable progress in realizing various regimes of coupling between the electromagnetic field and semiconductor quantum dots (QDs) has been achieved recently. Observations of phenomena related to the interaction of photons with discrete states in self-assembled dots such as ground state Rabi oscillations \cite{Zrenner}, weak \cite{Michler,Moreau,Happ} and strong \cite{Reithmaier,Yoshie} coupling regimes in various microcavity structures have strengthened the picture of the optical electron-hole pair excitation in a QD as a coherent two-level transition. QDs are considered as candidates for quantum information processing based on cavity quantum electrodynamics \cite{Imamoglu}. However, most of the work has concentrated on the absorptive coupling of QDs to the electromagnetic field. Almost no attention has been paid to the possibilities of dispersive coupling of QDs despite the fact that dispersive coupling has been exploited with single atoms in optical resonators to implement a conditional phase shift \cite{Turchette} and quantum-nondemolition measurements \cite{Brune}. In the light of this and along with the progress in the deterministic coupling of QDs to solid-state cavity devices \cite{Badolato} application of dispersive coupling schemes with QDs are very attractive possibilities.

In this letter we report simultaneous detection of forward and backward Rayleigh scattering of narrow-band laser light by a single QD. We employed a Fabry-P\'{e}rot setup at cryogenic temperature to detect and manipulate the interference of the laser field with its Rayleigh scattered component. Near the QD optical transition frequency the interference leads to resonances both in the differential transmission and reflectivity which were measured simultaneously. The method allows us to determine experimentally the dielectric polarizability function of a single self-assembled InAs/InGaAs QD. Our model describes the line shape evolution of the differential transmission and reflectivity spectra. We show that either optical signal can be used to deduce both the decoherence rate and the oscillator strength of the corresponding optical transition yielding its total scattering cross-section. Our results exemplify that confocal detection of reflectivity provides a high-resolution spectroscopy tool complementary to the near-field optics as used recently for natural QDs \cite{Lienau}. In the context of dispersive coupling schemes our work demonstrates that the dispersive nature of the QD is accessible experimentally.

\begin{figure}[t]
\includegraphics[scale=0.85]{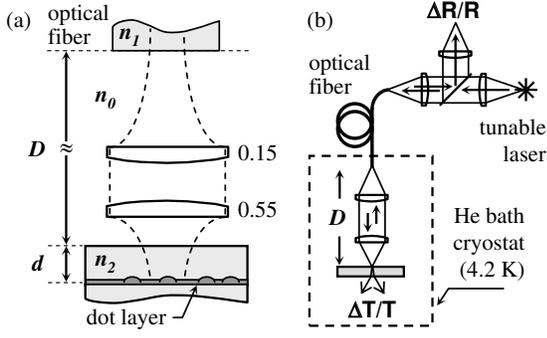}
\caption{\label{fig1}Setup schematics. (a) The polished fiber end and the sample surface separated by two aspheric lenses with numerical apertures 0.15 and 0.55 form a Fabry-P\'{e}rot cavity with a characteristic length of $D=10$~mm. The refractive indices of the fiber, sample and free space in between are $n_{1}$, $n_{2}$ and $n_{0}$, respectively. The QD layer is located $d=150$~nm below the semitransparent surface. (b) The excitation light of a tunable laser is delivered with an optical single-mode fiber to the interferometer in a helium bath cryostat (dashed box) at 4.2~K. Below the sample a photodiode detects the differential transmission $\m{\Delta T/T}$. The differential reflectivity $\m{\Delta R/R}$ is measured with a detector at room temperature.}
\end{figure}

The interferometry setup used in our experiments is depicted schematically in Fig.~\ref{fig1}(a). The planar Fabry-P\'{e}rot interferometer with a cavity length of $D=10$~mm is formed by the polished glass fiber end and the sample surface. The separation between the surface and the QD layer $d=150$~nm was determined by the sample growth sequence. The microscope objective consisting of two aspheric lenses with numerical apertures (NA) 0.15 and 0.55 aligns the cavity mirrors in a confocal arrangement. The optical path for interferometry detection is shown in Fig.~\ref{fig1}(b). The light of a tunable narrow-band laser is guided with a single-mode glass fiber to the interferometry setup which was cooled down to cryogenic temperature (4.2~K). The photodiode placed directly below the sample detects the QD forward scattering as differential transmission signal $\m{\Delta T/T}$ whereas the backward scattering is detected with a photodiode at room temperature which measures the differential reflectivity $\m{\Delta R/R}$.

\begin{figure}[t]
\includegraphics[scale=0.85]{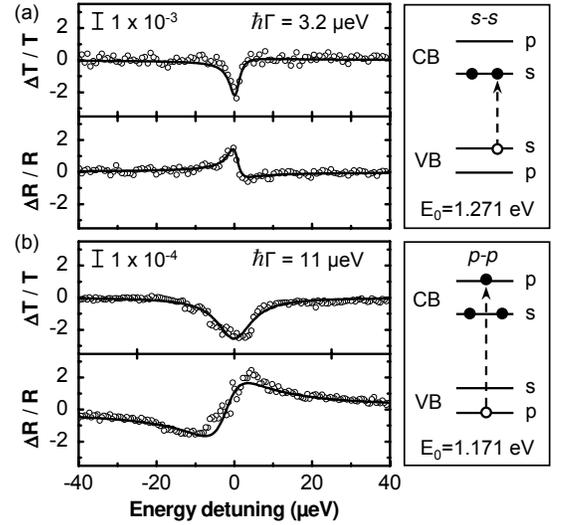}
\caption{\label{fig2} Differential transmission $\m{\Delta T/T}$ and reflectivity $\m{\Delta R/R}$ spectra of single QDs (open circles) and fits (solid lines) as described in the text. (a) Ground state valence band (VB) to conduction band (CB) excitation $s-s$ in a singly charged InGaAs QD. (b) Excited state transition $p-p$ in a single InAs dot with two resident electrons. Note different ordinate units in (a) and (b). The energy detuning was achieved through gate voltage induced Stark-shift.}
\end{figure}

Two different samples were investigated. InAs QDs \cite{Warburton1} in sample A were used to probe the excited state transition $p$- to $p$-shell (Fig.~\ref{fig2}, right panel) resonant with the laser energy E$_{0}=1.171$~eV ($\lambda=1058.9$~nm). For the ground state transition $s$- to $s$-shell resonant with E$_{0}=1.271$~eV ($\lambda=975.6$~nm), sample B containing InGaAs QDs \cite{Garcia} was used. In both samples, the self-assembled QDs were embedded in a field-effect structure. The QD layer was grown 25~nm above the back contact and 150~nm below the metalized gate on the sample surface with a 120~nm GaAs/AlAs superlattice in between. The device allows us to control precisely the number of resident electrons in an individual QD \cite{WarburtonN} and to exploit the Stark-shift for modulation spectroscopy \cite{Alen}.

\begin{figure}[t]
\includegraphics[scale=0.85]{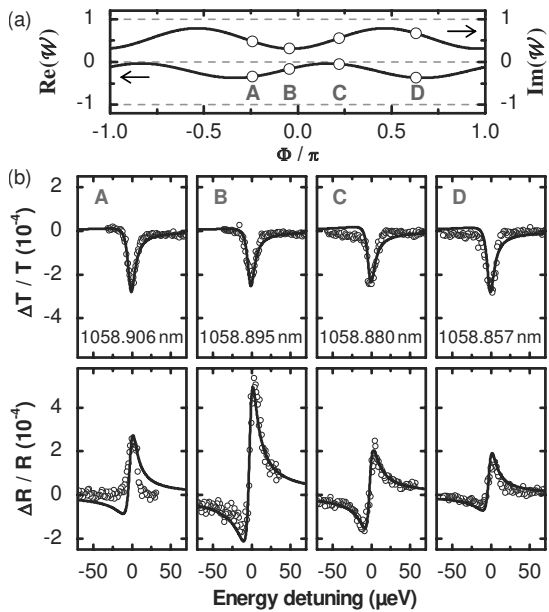}
\caption{ \label{fig3}(a) Absorptive and dispersive weighting function plotted against phase $\Phi$ in units of $\pi$. A, B, C and D mark the phase of the corresponding differential transmission and reflectivity spectra of a single InAs quantum dot represented in (b) as a function of detuning. Solid lines in (b) are fits to the data (open circles) according to the model described in the text.}
\end{figure}

The right panel of Fig.~\ref{fig2} depicts schematically the relevant optical interband transitions. The ground state excitation $s-s$ involves a QD charged with one resident electron such that only one resonance is present \cite{PRL}, the $p-p$ dipole transition is excited resonantly in a QD charged with two electrons \cite{Alen}. In the left panel of Fig.~\ref{fig2} typical differential transmission and reflectivity spectra of single QDs are shown. The linewidths of the $\m{\Delta T/T}$ spectra $\hbar \Gamma=3.2~\mu$eV and $\hbar \Gamma=11~\mu$eV measured at 4.2~K for the ground and the first excited state, respectively, are consistent with previously reported results \cite{Alen,PRL}. In the absence of interference effects, the transmission signature of a QD in a weakly focused Gaussian beam can be treated in the plane wave scattering limit \cite{Alen,Khaled}. The forward scattered resonance is a Lorentzian of purely absorptive character. The low finesse Fabry-P\'{e}rot cavity in the present experiment is expected to modify the transmission spectrum only moderately. This is confirmed in Fig.s~\ref{fig2} and \ref{fig3}. However, for the backward scattered light the tuning of the cavity plays a major role leading to a high sensitivity of the $\m{\Delta R/R}$ signal to interference effects (Fig.~\ref{fig3}).

In order to describe the line shape evolution of both the differential transmission and reflectivity in Fig.~\ref{fig3}, we model the following simplified situation: a QD excitonic dipole scatterer is positioned at distance $d$ on one side of an optical cavity of length $D$ with effective contrast $F=-4r_{1}r_{2}/(1+r_{1}r_{2})^2$ (effective finesse is $\mathcal{F}=\pi \sqrt{F}/2$) and corresponding characteristic Fabry-P\'{e}rot transmission function $\mathcal{T}(\Phi)=1/(1+F\sin^2\Phi)$. The effective amplitude reflectance coefficients of the fiber $r_{1}$ and the sample surface $r_{2}$ are determined through the effective refractive indices $n_{1}$ and $n_{2}$ as well as the vacuum refractive index $n_{0}=1$ through relations $r_{1}=(n_{1}-n_{0})/(n_{1}+n_{0})$ and $r_{2}=(n_{0}-n_{2})/(n_{0}+n_{2})$. Length scales $d$ and $D$ establish relative phase angles $\varphi= k_2 d$ between the QD response field and the laser field reflected at the sample-vacuum interface as well as $\Phi= k_{0} D$ between the QD field and the cavity field. Here, $k_2=2\pi n_2 /\lambda$ and $k_{0}=2\pi n_{0}/\lambda$ are the wave vectors in the embedding material and vacuum, respectively.

The complex polarizability $\tilde{\alpha}_{QD}$ of a QD two-level dipole transition driven by a harmonic laser field of frequency $\omega_{L}$ is a function of the decoherence rate $\gamma$ and laser detuning $\delta=\omega_{L}-\omega_{0}$ from the resonance frequency $\omega_{0}$. The decoherence rate is also linked to the full-width at half-maximum linewidth through $2 \hbar \gamma=\hbar \Gamma$. In resonance approximation with $\gamma\ll\omega_{0}$, as satisfied for QD transitions at cryogenic temperatures, the linear response function is given by:
\begin{eqnarray}
\label{eq:chi}
\nonumber && \tilde{\alpha}_{QD}  = - \tilde{\alpha}_{R}+i  \tilde{\alpha}_{I} \, ; \\
 && \tilde{\alpha}_{R}  =   \tilde{\alpha}_{0} \frac{\delta \gamma}{\delta^2+\gamma^2}\, , \, \tilde{\alpha}_{I} = \tilde{\alpha}_{0} \frac{\gamma^2}{\delta^2+\gamma^2}
\end{eqnarray}
\noindent The on-resonance maximum polarizability $\tilde{\alpha}_{0}=e^2 f /(2\gamma m_0 \varepsilon_0 \omega_0)$ is determined by the oscillator strength $f$ and the decoherence rate ($e$ is the electron charge, $m_0$ the free electron mass, and $\varepsilon_0$ the vacuum permittivity). $\tilde{\alpha}_{0}=\sigma_0/k_2$ can be interpreted as the scattering volume of a single QD, with the maximum total scattering cross-section $\sigma_0$ and the scattering depth $k_2$ of one wave number in the optical medium. On the other hand, for weakly focussed Gaussian beams the maximum contrast in an optical transmission experiment on resonance $\alpha_{0}=\sigma_0/\mathcal{A}$ is given by the ratio of the total scattering cross-section to the focal spot area $\mathcal{A}$ \cite{Focalspot}. The dispersive and absorptive responses of the dipole transition in Eq.~\ref{eq:chi} are identified with the real part $\tilde{\alpha}_{R}$ and the imaginary part $\tilde{\alpha}_{I}$, respectively.

In order to model the line shape evolution we derive analytical expressions for the differential transmission and reflectivity using the transfer-matrix method of light propagation \cite{Khaled}. With the definitions
\begin{eqnarray}
\label{eq:hw}
\mathcal{H}(\Phi) & = & -\frac{(1-r_{1}^2)(1-r_{2}^2)}{(r_{1}+r_{2})^2+(1+r_{1}r_{2})^2F\sin^2\Phi} \, \, ,\\
\nonumber \mathcal{W}(\varphi,\Phi) & = & \frac{(1+r_{1}^2)r_{2}e^{2i\varphi}+r_{1}[r_{2}^2e^{2i(\varphi-\Phi)}+e^{2i(\varphi+\Phi)}]}{(1+r_{1}r_{2})^2}
\end{eqnarray}
\noindent and the Fabry-P\'{e}rot transmission function $\mathcal{T}(\Phi)$, the differential transmission and reflectivity can be written as \cite{Khaled}:
\begin{eqnarray}
\label{eq:drr}
\nonumber \m{\Delta T/T}  = \frac{k_2}{\mathcal{A}}\{-\tilde{\alpha}_{I} +\mathcal{T} [\tilde{\alpha}_{I} \m{Re}(\mathcal{W}) -\tilde{\alpha}_{R}\m{Im}(\mathcal{W})]\} \\
\m{\Delta R/R}  =\frac{k_2}{\mathcal{A}}\{ \mathcal{H}\, \mathcal{T} [\tilde{\alpha}_{I}\m{Re}(\mathcal{W})-\tilde{\alpha}_{R}\m{Im}(\mathcal{W})]\}
\end{eqnarray}
\noindent Dependent on the phase angles $\varphi$ and $\Phi$, the function $\mathcal{W}$ weights the contributions of the absorptive and dispersive response as displayed in Fig.~\ref{fig3} for the case of resonant $p-p$ excitation. While $d$ and thus $\varphi$ is fixed by sample design, the phase $\Phi$ can be tuned either by the cavity length $D$ or the laser wavelength $\lambda$. We choose to tune the latter parameter which can be controlled very precisely. In the working point C the conditions for the optical reflectivity signal are such that the dispersive part is finite but the absorptive is nearly zero. The corresponding spectrum reveals a purely dispersive line shape and is depicted both in Fig.~\ref{fig2} and in the lower panel of Fig.~\ref{fig3} together with other spectra recorded at fixed phases marked A, B, and D. All fits to the data were obtained with Eq.~\ref{eq:drr}. The parameters are $\alpha_{0}=2\cdot 10^{-4}$ and $\hbar \Gamma=11~\mu$eV, resulting in an oscillator strength of $\m{f} \sim 10$, and $n_{1}=1.4$, $n_{2}=2.9$. The corresponding finesse of the Fabry-P\'{e}rot resonator is $\mathcal{F}\simeq 1$ comparable to 1.2 as calculated with refractive indices of $n_{1}=1.5$ for the glass fiber and $n_{2}=3.5$ for GaAs. The effective refractive indices $n_{1}$ and $n_{2}$ used in the model account for omitted superlattice, the metalized gate and potential cavity misalignment. The transmission is modified only slightly due to the low finesse of the cavity. The reflectivity spectra, however, reveal the full cyclic transition from an absorptive to a dispersive spectrum and back while the phase $\Phi$ of the interferometer is varied by $\pi$. The evolution is well reproduced by our model, although it overestimates slightly the asymmetry in the spectrum A. This notwithstanding, our results demonstrate clearly that the reflectivity signal can be modeled in detail to provide the oscillator strength and decoherence rate of the optical transition.

In summary, our work demonstrates that high-resolution interferometric Rayleigh scattering can be performed on a single self-assembled dot. The complex dielectric polarizability function of a single QD is accessible experimentally. Our results encourage the use of the reflectivity signal which can be modeled elaborately thus providing both the decoherence rate and the oscillator strength of the QD optical transition.

The authors would like to thank A. Imamo\u{g}lu and H. Weinfurter for helpful discussions. Financial support from SFB 631 (DFG), EPSRC (UK) and SANDiE (EU) is gratefully acknowledged. B. A. was supported by the Marie Curie European training program.


\begin{thebibliography}{}
\bibitem{Focalspot} $\mathcal{A}=1.13 \cdot$FWHM$^2$ for a Gaussian focal spot. In our experiments, the full-width at half-maximum is FWHM=1.3 $\mu$m.
\end{thebibliography}
\end{document}